# Voltage-induced thin-film superconductivity in high magnetic fields

Jabir Ali Ouassou,[1] Tom Doekle Vethaak,[1] and Jacob Linder[1]

[1]*Center for Quantum Spintronics, Department of Physics, Norwegian University of Science and Technology, NO-7491 Trondheim, Norway*

(Dated: July 10, 2018)

We predict that superconductivity in thin films can be stabilized in high magnetic fields if the superconductor is driven out of equilibrium by a DC voltage bias. For realistic material parameters and temperatures, we show that superconductivity is restored in fields many times larger than the Chandrasekhar–Clogston limit. After motivating the effect analytically, we perform rigorous numerical calculations to corroborate the findings, and present concrete experimental signatures. On the technical side, we also introduce a new form for the nonequilibrium kinetic equations, which generalizes and simplifies previous formulations of the problem.

*Introduction.*—It is well-known that magnetism is harmful to conventional superconductivity; the mechanisms responsible are a diamagnetic orbital effect and paramagnetic spin effect. The orbital effect refers to the Lorentz force felt by electrons moving in a magnetic field, which forces the electronic condensate to rotate. That requires kinetic energy, and eventually makes condensation unfavourable. This can be neglected in thin films with in-plane fields, since currents perpendicular to the plane are suppressed [1]. The spin effect refers to the magnetic spin-splitting of the electronic dispersion relation. Since a conventional Cooper pair consists of two electrons with opposite spins, this results in a momentum mismatch between the electrons in the pair. In clean systems, this may lead to an inhomogeneous superconducting state [2–4]. However, in dirty thin films, impurity and surface scattering prevent such an FFLO state from forming [2], and the spin effect just causes depairing instead. Superconductivity can therefore survive only up to the Chandrasekhar–Clogston limit $m = \Delta_0/\sqrt{2}$ [5, 6], where $m$ is the Zeeman-splitting of the magnetic field and $\Delta_0$ the zero-temperature gap of a bulk superconductor. In this paper, we show that this fundamental limit can be circumvented using a surprisingly simple trick: voltage-biasing the superconductor. The results are directly applicable to the dawning field of superconducting spintronics, where stabilizing superconductivity in proximity to magnetic elements is paramount [7–10].

Fig. 1 illustrates relevant experiments. The centerpiece is a thin-film superconductor exposed to an in-plane magnetic field. In theory, it does not matter whether this field is provided by a proximity effect or external source. However, a proximity effect provides a fixed field strength, limiting the parameter space one can explore with a single sample. On the other hand, externally inducing a Zeeman field $m \sim \Delta_0$ requires tens of teslas. Thus, the ideal solution is a combination: a ferromagnet produces a large offset $m = m_0 \sim \Delta_0$, while an external field tunes it to $m = m_0 + \mu_B H_{\text{ext}}$, where $\mu_B$ is the Bohr magneton. Finally, the superconductor is voltage-biased via tunneling contacts, providing an additional control parameter.

In Fig. 1(a), the voltage is applied across the superconductor. This both induces a nonequilibrium distribution there and injects a current that can be used as an observable. In Fig. 1(b), transverse wires are used to manipulate the distribution without any charge accumulation or current injection [11–13]. Herein, we focus on Fig. 1(a), and assume $\ell_e < \xi < L < \ell_{\text{in}}$ for an elastic mean free path $\ell_e$, diffusive coherence length $\xi$, system length $L$, and inelastic scattering length $\ell_{\text{in}}$.

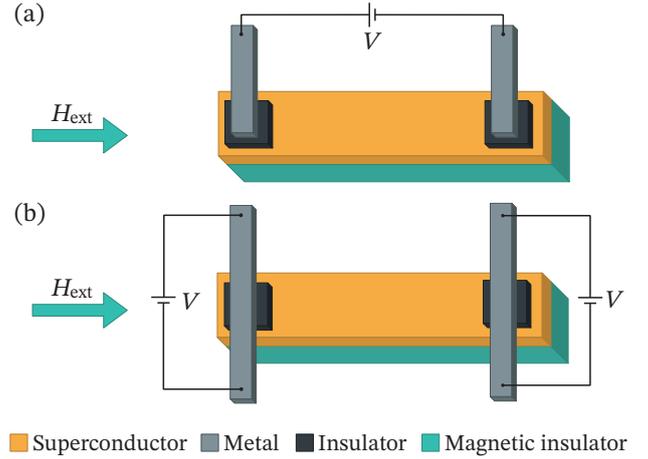

FIG. 1: Suggested experiments. A magnet induces a field $m = m_0$ in the superconductor, while an external field $H_{\text{ext}}$ shifts it to $m = m_0 + \mu_B H_{\text{ext}}$. The spin-split superconductor is then subjected to a voltage bias $V$, which stabilizes the superconducting state.

While both spin-split and voltage-biased superconductors have been investigated for a long time, a number of interesting discoveries have been made in recent years [14–27]. For instance, Bobkova and Bobkov [14] pointed out that there is a regime around the Chandrasekhar–Clogston limit where both a superconducting and normal state are allowed. This *bistability* means that if the field is varied adiabatically, and the metastable states relax slowly, one might observe a superconducting hysteresis effect. Snyman and Nazarov [15] had previously predicted the same kind of bistability in voltage-biased superconductors. Moreover, their results for the gap were similar to previous results for spin-split superconductors as discussed by Moor *et al.* [16], who showed that the gap equations for spin-split and voltage-biased superconductors are equivalent. Given the close analogy between these phenomena, a natural question is: what happens when both are present?

*Analytical motivation.*—Let us first consider a conventional superconductor in equilibrium, without any fields or voltages. The order parameter then satisfies a selfconsistency equation

$\Delta = N_0 \lambda F(\Delta)$, where the gap function $F(\Delta)$ is given by [28–30]

$$F(\Delta) = \frac{1}{2} \int_{-\omega_c}^{+\omega_c} d\epsilon \, \text{Re}[f(\epsilon)] \, h(\epsilon). \quad (1)$$

Here, $N_0$ is the density of states (DOS) at the Fermi level, $\lambda$ the BCS coupling constant, $f(\epsilon) = \Delta/\sqrt{\epsilon^2 - \Delta^2}$ the pair amplitude, $h(\epsilon) = \tanh(\epsilon/2T)$ the distribution function, $\epsilon$ the quasiparticle energy, $\omega_c$ the Debye cutoff, and $T$ the temperature. We measure energies relative to the zero-temperature gap $\Delta_0$, temperatures relative to the critical temperature $T_c$, and set the cutoff $\omega_c = 30\Delta_0$. Finally, in the weak-coupling limit the above are related by $N_0\lambda \approx 1/\log(2\omega_c/\Delta_0)$ and $\Delta_0 \approx \pi e^{-\gamma} T_c$, where $\gamma \approx 0.57722$ is the Euler–Mascheroni constant [28].

In response to a Zeeman-splitting field $m$, the quasiparticle energies become spin-split according to $\epsilon \to \epsilon \pm m$, and the pair amplitude in Eq. (1) therefore ends up taking the form $[f(\epsilon + m) + f(\epsilon - m)]/2$. On the other hand, when a voltage $V$ is applied over the superconductor, the electronic distribution functions of the contacts are shifted to $h(\epsilon \pm eV/2)$. By linear combination of the electron and hole distributions, this can be decomposed into a charge mode $[h(\epsilon \pm eV/2) - h(\epsilon \mp eV/2)]/2$ and energy mode $[h(\epsilon \pm eV/2) + h(\epsilon \mp eV/2)]/2$ [13, 29, 31]. The charge mode, which is related to charge accumulation, relaxes quickly inside the superconductor [17]. The energy mode, on the other hand, remains constant throughout the superconductor, and couples to the order parameter in Eq. (1) instead of $h(\epsilon)$. For further discussion of the relevant modes, see the supplemental. In reality, voltage-biasing the superconductor also induces a supercurrent, which manifests as a phase-winding of the pair amplitude $f(\epsilon)$ and a suppression of the gap. In the tunneling limit, this phase-winding is small enough to be neglected in the selfconsistency equation. However, when we later in this paper study the system fully numerically, we also take phase-winding and nonequilibrium spin modes into account. Making the above modifications to Eq. (1), we find that for a magnetic field $m$ and voltage $V$ the gap function becomes

$$\begin{aligned} F(\Delta, m, eV/2) &= \frac{1}{8} \int_{-\omega_c}^{+\omega_c} d\epsilon \, \text{Re}[f(\epsilon - m) + f(\epsilon + m)] \, h(\epsilon + eV/2) \\ &+ \frac{1}{8} \int_{-\omega_c}^{+\omega_c} d\epsilon \, \text{Re}[f(\epsilon - m) + f(\epsilon + m)] \, h(\epsilon - eV/2). \end{aligned} \quad (2)$$

Following the same approach as Moor et al. [16], we substitute $\epsilon' \equiv \epsilon \pm eV/2$ into the above to express the voltages as equivalent magnetic fields. Formally, the integration limits have to be shifted accordingly—but since $\omega_c \gg \Delta, m, eV/2$, this is inconsequential. After some reordering, the result becomes

$$\begin{aligned} F(\Delta, m, eV/2) &= F(\Delta, m - eV/2, 0)/2 \\ &+ F(\Delta, m + eV/2, 0)/2. \end{aligned} \quad (3)$$

At this point, we find three properties worth remarking. Firstly, the right-hand side is invariant under $eV/2 \leftrightarrow m$. In other words, the gap responds in precisely the same way to an applied voltage and magnetic field. Moreover, plotted as a function of these two control parameters, the superconducting gap should be symmetric around the diagonals $eV/2 = \pm m$. Secondly, in spin-split superconductors without any voltage, it is known that a superconducting solution $\Delta = \Delta_0$ exists as long as the magnetic field $m < \Delta_0$. Applied to Eq. (3) above, $\Delta = \Delta_0$ should remain a valid solution for $|m| + |eV/2| < \Delta_0$. Parts of this regime is bistable, and admits a normal-metal solution as well. Finally, we note that the effects of a voltage and magnetic field cancel in the first term, but act constructively in the second. This becomes especially clear if we tune the voltage to $eV/2 = m$, where we find the peculiar result:

$$F(\Delta, m, m) = [F(\Delta, 0, 0) + F(\Delta, 2m, 0)]/2. \quad (4)$$

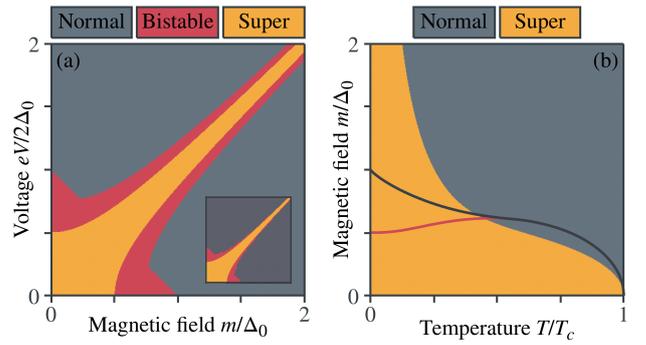

FIG. 2: Analytically calculated phase diagrams for a voltage-biased spin-split superconductor at (a) $T = 0$ and (b) $eV/2 = m$. The inset shows numerical results, which qualitatively match the analytical ones. For comparison, we overlaid phase transition lines for $V = 0$ in (b), where red marks the onset of bistability, and black where superconductivity vanishes entirely.

The first term is just the gap function in the absence of fields and voltages, which by itself always results in superconductivity at low temperatures. The second is the gap function for a superconductor with a magnetic field $2m$, which only contributes to superconductivity until $m = \Delta_0/2$. We therefore expect the combination to yield a bulk gap until $m = \Delta_0/2$, but also produce a weaker superconducting solution for much higher fields, since half the gap function is independent of $m$.

In Fig. 2, we show phase diagrams for the voltage-biased spin-split superconductor, which were calculated using Eq. (2). For more details on how the phases were classified, and in particular how bistability was checked, see the supplemental. Fig. 2(a) demonstrates all the qualitative features motivated above: superconducting solutions exist for $eV/2 \approx m$ and $|eV/2| + |m| < \Delta_0$, and are symmetric around $eV/2 = m$. For $eV/2 \approx m$, the system is not even bistable, but *only* admits superconducting solutions. In other words, no stable normal-state solution exists along $eV/2 \approx m$ at low temperatures, even for fields much larger than the Chandrasekhar–Clogston limit. For high magnetic fields $m > \Delta_0$, the effect is particularly striking: there is no superconductivity in the system *until* a voltage $eV/2 \approx m$ is applied. *In other words, a voltage bias* *can be used to allow coexistence of superconductivity and a Zeeman-splitting magnetic field that is much larger than both the Chandrasekhar–Clogston limit and the bulk gap.*

Fig. 2(b) shows the temperature-dependence for $eV/2 = m$. This phenomenon is clearly a low-temperature effect: the limiting magnetic field seems to diverge as $T \to 0$, while a comparison with the overlays indicate that the cancellation becomes insignificant after $\sim 0.4 T_c$. However, the plot also shows that superconductivity is stabilized in high fields $m > \Delta_0$ at temperatures up to $T \approx 0.25 T_c$, which for e.g. niobium corresponds to reasonable value $\sim 2.3$ K. Going down to $\sim 1$ K, one can even expect superconductivity for $m > 2\Delta_0$.

Note that the stabilization of superconductivity for $eV/2 = m$ is similar to an effect previously reported by Bobkova and Bobkov [26, 27]. They found that if one applies a spin-dependent voltage $eV_\uparrow/2 = -eV_\downarrow/2 = m$, which can be achieved using voltage-biased half-metallic contacts, superconductivity is recovered. Our effect is however qualitatively different, since it arises for a spin-independent quasiparticle distribution and purely electric voltage bias using normal metal contacts. Related effects were demonstrated in Refs. [32–34], which found that the critical current in Josephson junctions behaved symmetrically with respect to voltages and magnetic fields.

*Numerical treatment.*— We use the quasiclassical Keldysh formalism [13, 22, 29, 35, 36], where observables are described via an $8 \times 8$ propagator in Keldysh $\otimes$ Nambu $\otimes$ Spin space,

$$\check{g} = \begin{pmatrix} \hat{g}^R & \hat{g}^A \\ 0 & \hat{g}^K \end{pmatrix}. \tag{5}$$

The components are related by the identities $\hat{g}^K = \hat{g}^R \hat{h} - \hat{h} \hat{g}^A$ and $\hat{g}^A = -\hat{\tau}_3 \hat{g}^{R\dagger} \hat{\tau}_3$, where $\hat{h}$ is the distribution matrix, and $\hat{\tau}_3 = \text{diag}(+1, +1, -1, -1)$ is a Pauli matrix in Nambu space. It is therefore sufficient to determine the retarded propagator $\hat{g}^R$ and distribution matrix $\hat{h}$. It is commonly stated that $\hat{g}^R$ describes the equilibrium state, while $\hat{h}$ describes the nonequilibrium one. However, this is actually incorrect for a superconductor, since $\hat{g}^R$ implicitly depends on $\hat{h}$ via the selfconsistently determined gap $\Delta$. In practice, one therefore has to alternate between solving a diffusion equation for $\hat{g}^R$, a kinetic equation for $\hat{h}$, and a selfconsistency equation for $\Delta$, until all three converge.

The propagator is governed by the Usadel equation [36],

$$i\xi^2 \nabla(\check{g} \nabla \check{g}) = [\hat{\Delta} + \epsilon \hat{\tau}_3 + m \hat{\sigma}_3, \check{g}]/\Delta_0, \tag{6}$$

where $\hat{\Delta} = \text{antidiag}(+\Delta, -\Delta, +\Delta^*, -\Delta^*)$ is the gap matrix, $\hat{\sigma}_3 = \text{diag}(+1, -1, +1, -1)$ a Pauli matrix in spin space. The film is voltage-biased via tunneling contacts, which we model with Kupriyanov–Lukichev boundary conditions [37] to normal reservoirs with chemical potentials $\mu = \pm eV/2$ relative to the superconductor. The interfaces are characterized by the ratio of tunneling to bulk conductance, which we set to a moderate value $G_T/G_0 = 0.3$, which for an $8\xi$ thick Nb superconductor corresponds to an average channel transparency $\sim 1\%$. We accounted for inelastic scattering using the Dynes approximation $\epsilon \to \epsilon + 0.01 i \Delta_0$ [38, 39]. Finally, we set the superconductor length $L = 8\xi$; in general, we found a stronger recovery of superconductivity for longer junctions, but at $L = 8\xi$ the gap had nearly saturated. We also performed a number of tests using transparent interfaces, and found similar results as long as the superconductor was made sufficiently long; this indicates that the results are not very sensitive to the specific material parameters used. The retarded propagator $\hat{g}^R$ was Riccati-parametrized for stability [40], and solved for in the same way as usual [41]. The gap function can be written as an integral of the singlet anomalous component of the Keldysh propagator $\hat{g}^K$ [28]. Physical observables, such as the current and DOS, were calculated from the quasiclassical propagators using standard formulas [9, 13, 22, 29, 35].

As for the kinetic equations, we have derived a new form which generalizes and simplifies previous results. Our approach is similar to the treatment of nonequilibrium S/N systems in Refs. [13, 29], and especially the treatment of S/F systems with spin-flip and spin-orbit scattering in Refs. [19–22]. However, we extend their results to an Usadel equation with a completely general second-order self-energy $\nabla(\check{g} \nabla \check{g}) \sim [\hat{\Sigma}^{(1)} + \hat{\Sigma}^{(2)} \check{g} \hat{\Sigma}^{(2)}, \check{g}]$, and derive accompanying boundary conditions for strongly polarized magnetic interfaces based on Ref. [42]. We do not make any simplifying assumptions, so the results can be used for systems with voltages, spin-voltages, temperature gradients, spin-temperature gradients, and any combination of spin projections. The final result is an *explicit linear second-order differential equation* that takes a simple form,

$$\begin{aligned} M_{nm} \nabla^2 h_m = &- (\nabla M_{nm} + \boldsymbol{Q}_{nm}) \cdot \nabla h_m \\ &- (\nabla \cdot \boldsymbol{Q}_{nm} + V_{nm} + W_{nm}) h_m, \end{aligned} \tag{7}$$

where we implicitly sum over repeated indices. The distribution is parametrized as an 8-element vector $h$, which describes all charge, spin, heat, and spin-heat degrees of freedom. The coefficients $M$, $\boldsymbol{Q}$, $V$, $W$ are $8 \times 8$ matrices that depend on the retarded propagator $\hat{g}^R$ and self-energy factors $\hat{\Sigma}^{(1)}, \hat{\Sigma}^{(2)}$, but not the distribution $h$. In addition to being simple and general, this formulation is very efficient numerically since all coefficients are independent of $h$; in fact, it takes less time to solve than the Riccati-parametrized equations for $\hat{g}^R$. For more details about the kinetic equations, see the supplemental.

*Numerical results.*— Fig. 3 highlights some properties of the configuration $eV/2 = m$ that stabilizes superconductivity. Fig. 3(a) compares the analytical and numerical approaches: even though the former neglects both inelastic scattering and spatial variations, we find an impeccable agreement at low temperatures. Beyond the point $m = \Delta_0/2$, we see that the gap suddenly starts to decrease with $m$. At finite temperatures, we find that superconductivity remains until $m \approx 2\Delta_0$ at $T = 0.1 T_c$ and $m \approx \Delta_0$ at $T = 0.2 T_c$, in agreement with Fig. 2(b).

Fig. 3(b) shows the DOS in the center of the superconductor, which can be observed by scanning tunneling microscopy (STM). These predictions are interesting: at $m \approx \Delta_0/2$, *a gigantic zero-energy peak develops throughout the superconductor without destroying the singlet condensate.* For $m \gg \Delta_0$, another unusual state develops, manifesting as two half-filled BCS gaps far from the Fermi level $\epsilon = 0$. However, these are not unreasonable results: it is exactly what one would expect from a BCS DOS $N(\epsilon) = N_0 \, \text{Re}\left[|\epsilon|/\sqrt{\epsilon^2 - \Delta^2}\right]$, if one uses the gaps $\Delta$ in Fig. 3(a) and introduces spin-splitting $\epsilon \to \epsilon \pm m$. For $m \approx \Delta_0/2 \approx \Delta(m = \Delta_0/2)$, this results in two BCS-shapes that are shifted so that their coherence peaks overlap at $\epsilon = 0$, causing a zero-energy peak to manifest. At higher fields $m \gg \Delta_0$, we instead see two disjoint BCS-shapes. The spin-resolved DOS (not shown) confirms that there is actually a full spectral gap in the spin-down DOS at $\epsilon = +m$ and spin-up DOS at $\epsilon = -m$, causing the spin-independent DOS shown in Fig. 3(b) to exhibit two apparently half-filled spectral gaps.

In Fig. 4, we present another experimental signature. Fig. 4(a) shows that for a fixed field $m = \Delta_0$, no superconductivity exists without a voltage bias. At $eV/2 \approx \Delta_0$, superconductivity is suddenly stabilized; taking the superconductor to be e.g. niobium, the gap is then restored to $\Delta \approx 0.36$ meV at $T \approx 1$ K and $\Delta \approx 0.22$ meV at $T \approx 2$ K. This manifests as a spike in the otherwise ohmic current flowing through the junction, causing an excess current of $\sim 5\%$ at 1 K and $\sim 1\%$ at 2 K. Fig. 4(b) demonstrates that the same qualitative behaviour is expected for a fixed voltage $eV/2 = \Delta_0$ and varying magnetic field. This shows that there is similarly a regime where an applied magnetic field is required to induce superconductivity. Although Fig. 2 shows that the stable superconducting regime $eV/2 \approx m$ should be padded by a bistable regime, this bistabile regime shrinks considerably at finite inelastic scattering and temperature. So while we do find bistability numerically as well for very low temperatures $T = 0.01T_c$, the bistable regime is almost nonexistent for the parameters used in Fig. 4.

*Conclusion.*—We have shown that superconductivity can coexist with a Zeeman-splitting magnetic field far beyond the Chandrasekhar–Clogston limit $m = \Delta_0/\sqrt{2}$ if the superconductor is exposed to a voltage bias $eV/2 = m$. We present concrete setups for observing this effect in Fig. 1, and provide two experimental signatures: the peculiar spin-split DOS in Fig. 3, which can be measured using an STM; and the excess current in Fig. 4, which produces a significant deviation from the otherwise ohmic behaviour. If we take the superconductor to be niobium, the signals are very strong at 1 K, but can be observed at $m = \Delta_0$ for temperatures up to 2 K.

Possibilities for future work include investigating how robust this superconducting state is with respect to spin-flip scattering, spin-orbit scattering, and orbital depairing. It would also be interesting to check if a similar effect exists for unconventional high-temperature superconductors. Finally, our setup might be used as a novel circuit element for superconducting spintronic junctions. For instance, the $m \approx \Delta_0/2$ curve in Fig. 3(b) shows a gigantic zero-energy peak inside the superconductor, which is reminiscent of an intrinsic odd-frequency superconductor. An even more peculiar behaviour might arise for $m \gg \Delta_0$, when the spectral gaps of the spin bands do not overlap.

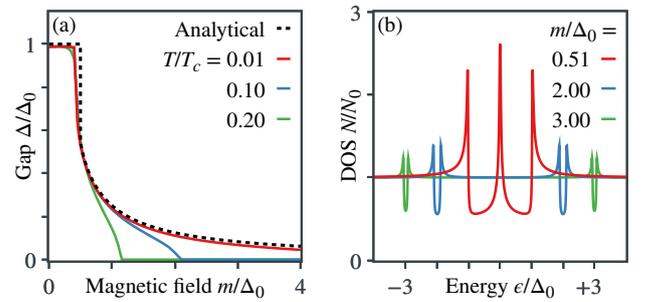

FIG. 3: We set $eV/2 = m$, and investigate how $m$ affects (a) the order parameter (b) the DOS at $T = 0.01T_c$. The dashed line shows Eq. (2) at $T = 0$, colored lines are finite-temperature numerical results.

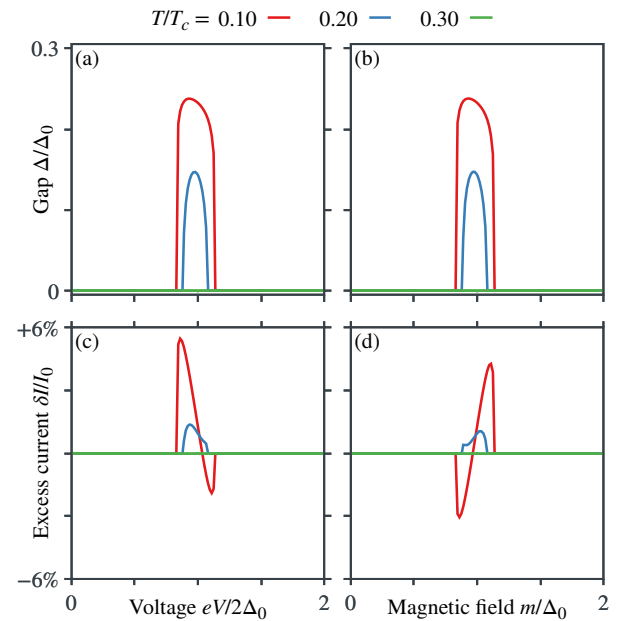

FIG. 4: Numerically calculated gap $\Delta$ for (a) fixed field $m = \Delta_0$ and varying voltage, (b) fixed voltage $eV/2 = \Delta_0$ and varying field. The temperatures $T$ are given in the legends above. Figs. (c–d) show corresponding deviations $\delta I$ from the normal-state current $I = GV$ due to superconductivity, normalized to the current $I_0$ at $eV/2 = \Delta_0$.

*Acknowledgments.*—We thank A. Golubov, W. Belzig, and M. Aprili for helpful discussions. The numerics was performed on resources provided by UNINETT Sigma2—the national infrastructure for high performance computing and data storage in Norway. This work was supported by the Research Council of Norway through grant 240806, and through its Centres of Excellence funding scheme grant 262633 "*QuSpin*".

# Supplemental information

Jabir Ali Ouassou,[1] Tom Doekle Vethaak,[1] and Jacob Linder[1]

[1]*Center for Quantum Spintronics, Department of Physics, Norwegian University of Science and Technology, NO-7491 Trondheim, Norway*

In Section I, we provide some additional details on how the phase diagrams in the main manuscript were calculated. We then derive a new formulation of the nonequilibrium kinetic equations in Section II. The resulting equations are completely general, computationally efficient, and yet straight-forward to implement numerically. Corresponding boundary conditions for magnetic interfaces to general nonequilibrium reservoirs are derived in Section III. We then show numerical results for the distribution in IV, and discuss the conversion to supercurrents in V.

## I. SOLVING THE SELFCONSISTENCY RELATION

The selfconsistency relation for the superconducting gap can be written $\Delta \sim F(\Delta, m, eV/2)$ in the presence of a magnetic field $m$ and voltage bias $V$, as discussed in the main manuscript. The simplest way to solve such equations is by fixpoint iteration: for each field $m$ and voltage $V$, one chooses an initial guess $\Delta = \Delta_1$, and calculates successive values using $\Delta_{n+1} \sim F(\Delta_n, m, eV/2)$. This is repeated until the difference $|\Delta_{n+1} - \Delta_n|$ between iterations drops below some acceptance threshold, at which point the system is said to have converged to a fixed point for the gap. These fixed points that the system converges toward correspond to minima in the free energy; this is not straight-forward to verify within the Usadel formalism, but can be found by comparison to the Ginzburg–Landau [1] and Bogoliubov–de Gennes [2] formalisms. Alternatively, it can be argued more heuristically that the Usadel equation should be possible to derive by minimizing some free energy—and its selfconsistent solution should converge towards these minima.

In many cases, the magnitude of the gap converges towards the same fixed point for any finite initial guess $\Delta \neq 0$. This fixed point is then called a *stable solution*, since the system converges back to the same point after perturbations. Some care must be taken with the normal-state solution $\Delta = 0$: mathematically, this is *always* a solution to the selfconsistency equation. However, below the critical temperature of the superconductor, it actually corresponds to a *local maximum* in the free energy. In this case, one finds that even infinitesimal perturbations of the initial state results in a divergence away from this point, which is why it is called an *unstable solution*. These solutions are not very interesting from a physical point-of-view, and can be discarded.

However, in some systems, the situation is more complicated. In a spin-split superconductor, there is a parameter regime $\Delta_0/2 < m < \Delta_0$ where the gap converges to a superconducting solution $\Delta = \Delta_0$ for large initial guesses, but a normal-state solution $\Delta = 0$ for small guesses. Both solutions are *locally stable* in the sense that they are robust to small perturbations, and correspond to different *local minima* of the free energy [1]. These two minima are separated by an energy barrier, which manifests as an unstable solution $\Delta = \Delta_0 \sqrt{2m/\Delta_0 - 1}$ where the free energy of the system is maximized [3]. In a voltage-biased superconductor, the exact same situation occurs for the voltages $\Delta_0/2 < eV/2 < \Delta_0$ [4]. Other situations where multiple locally stable solutions can exist include optically pumped systems [5], complex Josephson junctions [6], and supercooled type-I superconductors in a magnetic field [7].

Originally, this *bistability* was resolved by comparing the energies of the two minima, since the system should eventually relax to the global minimum. In *e.g.* the magnetic case, this yields the Chandrasekhar–Clogston limit $m = \Delta_0/\sqrt{2}$ as the exact transition point in the interval $\Delta_0/2 < m < \Delta_0$, where a first-order phase transition takes place [8, 9]. However, if the magnetic field is varied adiabatically beyond this point, the superconductor can in principle remain in a local minimum for some time before collapsing to the global minimum. Thus, one might observe a kind of superconducting hysteresis effect in this regime, and a more accurate characterization might be to call it bistable or hysteretic [1, 4]. In this manuscript, we take this view, and therefore classify the phase diagram of the junction into *superconducting*, *bistable*, and *normal* regions, where the bistable one might exhibit either a superconducting hysteresis or first-order phase transition depending on the relaxation times of the metastable states. Since it is not straight-forward to accurately calculate the free energy itself within the Usadel formalism, we do not calculate the thermodynamic transition lines, but these can be assumed to lie in the bistable regime.

After introducing the terminology, we now demonstrate how the phase diagram itself was calculated. In Fig. S1(a–c), we visualize how the superconducting gap $\Delta$ changes depending on the initial guess. Fig. S1(a) in particular visualizes the spin-split superconductor discussed above. For $m < \Delta_0/2$, the gap increases for $\Delta < \Delta_0$, decreases for $\Delta > \Delta_0$, and always converges to $\Delta = \Delta_0$. This is a stable superconducting regime. Conversely, for $m > \Delta_0$, the gap decreases to $\Delta = 0$ regardless of our initial guess. This is a stable normal-state solution. But for the intermediate regime $\Delta_0/2 < m < \Delta_0$, there are three distinct solutions for the gap [3]: a superconducting one $\Delta = \Delta_0$, a normal one $\Delta = 0$, and an unstable one inbetween. This is an example of the bistability discussed above. Fig. S1(b) shows the corresponding case for a voltage-biased superconductor, which behaves identically [4, 10]. Note that in Fig. S1(c), we also find a brief bistability between two superconducting solutions at $m = eV/2 \approx \Delta_0/2$; such regions were classified as *superconducting* and not *bistable* in the main manuscript.

Fig. S1(d–f) displays how the superconducting states were classified, using similar colors to Fig. 2 in the main manuscript. In practice, two different initial guesses $\Delta = 10^{-4}\Delta_0$ and $\Delta = 1.01\Delta_0$ are sufficient to identify both solutions in bistable regimes; this was done for $400 \times 400$ values of $m$ and $eV/2$ to construct Fig. 2. We note that our Fig. S1(d) is in agreement with Ref. 3, and Fig. S1(e) is in agreement with Ref. 4, while Fig. S1(f) represents a new result obtained in this paper.

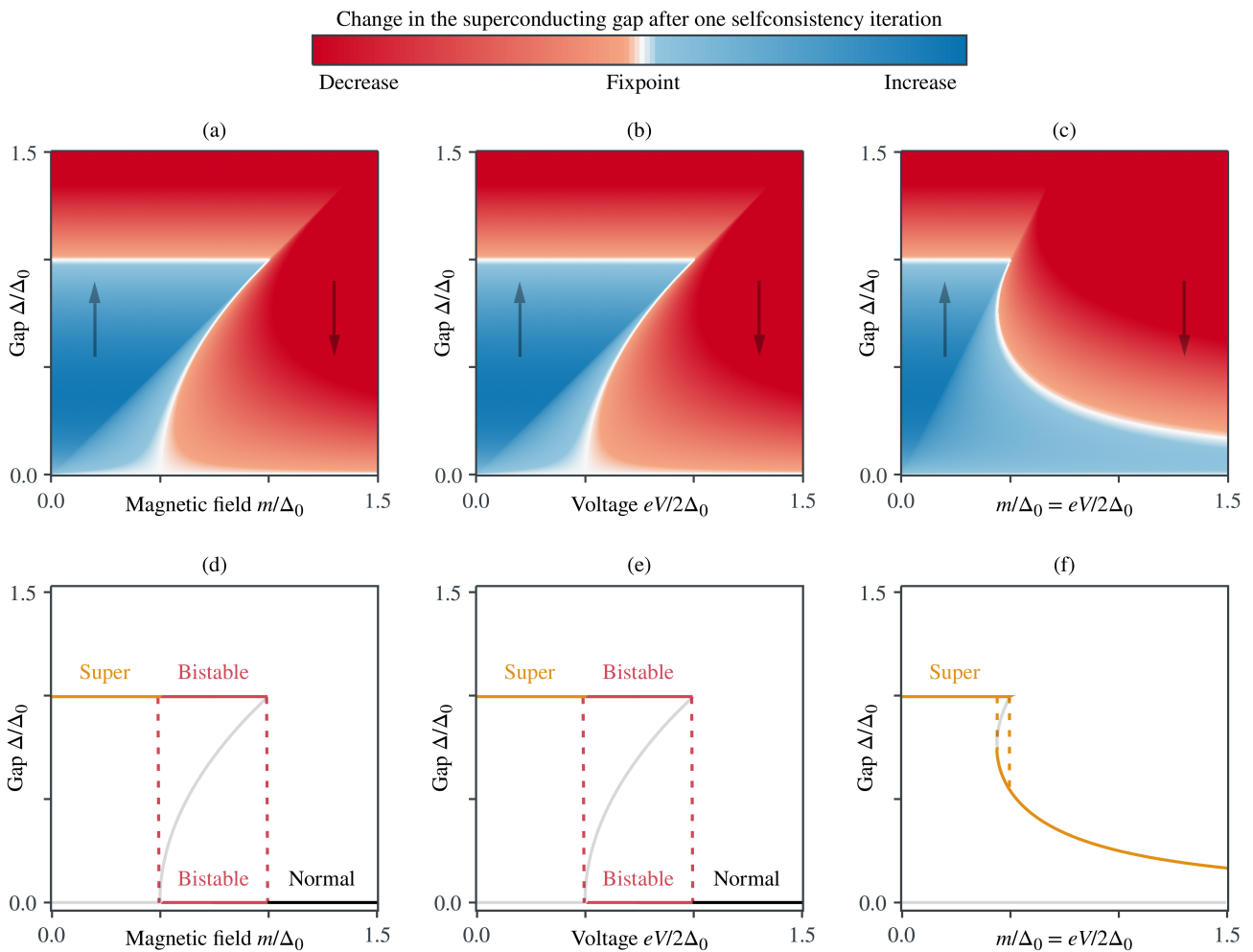

FIG. S1: Flow of the superconducting gap $\Delta$ between selfconsistency iterations for (a) varying magnetic field $m$ but zero voltage $V = 0$, (b) varying voltage $V$ but zero magnetic field $m = 0$, and (c) matched magnetic field and voltage $m = eV/2$. The gap spontaneously *increases* in blue regions, *decreases* in red regions, and comes to a standstill in white regions. Panels (d–f) show the inferred solutions for the gap, classified as superconducting (yellow), bistable (red), normal (black), and unstable (grey), *i.e.* using a colorscheme similar to Fig. 2 in the main manuscript.

## II. KINETIC EQUATIONS

In this section, we derive an *explicit linear ordinary differential equation* for the distribution function $\hat{h}$. The result is a highly computationally efficient form of the kinetic equations, which is also relatively straight-forward to implement numerically.

The starting point is the Usadel equation [11, 12], which describes diffusive materials in the quasiclassical limit. In terms of the matrix current $\check{I} \equiv D(\check{g}\nabla\check{g})$ and self-energy $\check{\Sigma}$, the Usadel equation can be written $\nabla \cdot \check{I} = -i[\check{\Sigma}, \check{g}]$ [13, 14]. However, for our purposes, we only require the Keldysh component:

$$\nabla \cdot \hat{I}^K = -i[\check{\Sigma}, \check{g}]^K. \quad (S1)$$

As we will see later, it will prove prudent to introduce a set of basis matrices $\hat{\rho}_n$ that span block-diagonal spin-Nambu space,

$$\hat{\rho}_0 \equiv \hat{\tau}_0 \hat{\sigma}_0, \quad \hat{\rho}_1 \equiv \hat{\tau}_0 \hat{\sigma}_1, \quad \hat{\rho}_2 \equiv \hat{\tau}_0 \hat{\sigma}_2, \quad \hat{\rho}_3 \equiv \hat{\tau}_0 \hat{\sigma}_3; \quad (S2)$$
$$\hat{\rho}_4 \equiv \hat{\tau}_3 \hat{\sigma}_0, \quad \hat{\rho}_5 \equiv \hat{\tau}_3 \hat{\sigma}_1, \quad \hat{\rho}_6 \equiv \hat{\tau}_3 \hat{\sigma}_2, \quad \hat{\rho}_7 \equiv \hat{\tau}_3 \hat{\sigma}_3. \quad (S3)$$

Here, $\hat{\tau}_0 \equiv \text{diag}(+\sigma_0, +\sigma_0)$ and $\hat{\tau}_3 \equiv \text{diag}(+\sigma_0, -\sigma_0)$ are the diagonal basis matrices in Nambu space, while $\hat{\sigma}_i \equiv \text{diag}(\sigma_i, \sigma_i^*)$ forms a complete basis for the spin structure, where $\sigma_i$ are the Pauli matrices. This lets us rewrite the distribution matrix $\hat{h}$ as

$$\hat{h} = h_n \hat{\rho}_n, \quad (S4)$$

where we use the summation convention on the right-hand side, and define the coefficients $h_n$ as traces with the basis matrices,

$$h_n \equiv \frac{1}{4} \text{Tr}[\hat{\rho}_n \hat{h}]. \quad (S5)$$

The kinetic equations take a simple form when written in terms of $h_n$ instead of $\hat{h}$, while Eq. (S4) makes it trivial to reconstruct the matrix structure afterwards. When implementing our results numerically, $h_n$ is treated as a real-valued 8-vector, while the kinetic equations will involve $8 \times 8$-matrices operating on it.

## A. Decomposition of the current

Combining the definitions $\check{I} = D(\check{g}\nabla\check{g})$ and $\hat{g}^K = \hat{g}^R\hat{h} - \hat{h}\hat{g}^A$, we find that the Keldysh component of the matrix current is

$$\hat{I}^K = D[(\hat{g}^R \nabla \hat{g}^R)\hat{h} - \hat{h}(\hat{g}^A \nabla \hat{g}^A)] \\ + D[(\nabla \hat{h}) - \hat{g}^R(\nabla \hat{h})\hat{g}^A]. \quad (S6)$$

Substituting in the parametrization $\hat{h} = h_m \hat{\rho}_m$, we see that the coefficients $h_m$ and $\nabla h_m$ can be factored out of the brackets,

$$\hat{I}^K = D[(\hat{g}^R \nabla \hat{g}^R)\hat{\rho}_m - \hat{\rho}_m(\hat{g}^A \nabla \hat{g}^A)] h_m \\ + D[\hat{\rho}_m - \hat{g}^R \hat{\rho}_m \hat{g}^A] \nabla h_m. \quad (S7)$$

If we then multiply the entire equation by $\hat{\rho}_n/4$ from the left and take the trace, the resulting equation can be written

$$I_n = Q_{nm} h_m + M_{nm} \nabla h_m, \quad (S8)$$

where we have defined the quantities

$$I_n \equiv \frac{1}{4} \mathrm{Tr}\left[\hat{\rho}_n \hat{I}^K\right], \quad (S9)$$

$$Q_{nm} \equiv \frac{D}{4} \mathrm{Tr}\left[\hat{\rho}_m \hat{\rho}_n (\hat{g}^R \nabla \hat{g}^R) - \hat{\rho}_n \hat{\rho}_m (\hat{g}^A \nabla \hat{g}^A)]\right], \quad (S10)$$

$$M_{nm} \equiv \frac{D}{4} \mathrm{Tr}\left[\hat{\rho}_n \hat{\rho}_m - \hat{\rho}_n \hat{g}^R \hat{\rho}_m \hat{g}^A\right]. \quad (S11)$$

This has a straight-forward physical interpretation [15–18]. The traces of $\hat{\rho}_n \hat{I}^K$ are proportional to the spectral charge, spin, heat, and spin-heat currents, meaning that $I_n$ describes the physically observable currents in the system. The right-hand side of Eq. (S8) then relates this to the distribution function $h_m$ and its derivative $\nabla h_m$. The term proportional to $h_m$ can be nonzero even in equilibrium, which means that $Q_{nm}$ can be identified as the supercurrent contribution. On the other hand, the term proportional to $\nabla h_m$ requires an inhomogeneous distribution function, so $M_{nm}$ is a resistive contribution.

If we now go back to the Usadel equation, and multiply that by $\hat{\rho}_n/4$ from the left and take the trace, we find the equation:

$$\nabla \cdot I_n = -\frac{i}{4} \mathrm{Tr}\left\{\hat{\rho}_n [\check{\Sigma}, \check{g}]^K\right\}. \quad (S12)$$

This will later be combined with Eq. (S8) to derive the kinetic equations. First, however, we need to express the right-hand side of the equation in terms of the distribution function $h_m$.

## B. First-order self-energy terms

When describing phenomena such as superconductivity and ferromagnetism, the self-energy matrix $\check{\Sigma} = \hat{\Sigma}$ is diagonal in Keldysh space and independent of the propagator $\check{g}$. This simplifies the commutator on the right-hand side of Eq. (S12):

$$[\check{\Sigma}, \check{g}]^K = [\hat{\Sigma}, \hat{g}^K]. \quad (S13)$$

Substituting in $\hat{g}^K = \hat{g}^R \hat{h} - \hat{h}\hat{g}^A$ and $\hat{h} = h_m \hat{\rho}_m$, we then find:

$$[\check{\Sigma}, \check{g}]^K = [\hat{\Sigma}, \hat{g}^R \hat{\rho}_m - \hat{\rho}_m \hat{g}^A] h_m. \quad (S14)$$

Going back to Eq. (S12), we therefore find that it can be written

$$\nabla \cdot I_n = -V_{nm} h_m, \quad (S15)$$

where we have defined the new quantity

$$V_{nm} \equiv \frac{i}{4} \mathrm{Tr}\left\{\hat{\rho}_n [\hat{\Sigma}, \hat{g}^R \hat{\rho}_m - \hat{\rho}_m \hat{g}^A]\right\}. \quad (S16)$$

Finally, we note that using the cyclic property of the trace, the above can be rewritten in the alternative form

$$V_{nm} = \frac{i}{4} \mathrm{Tr}\left\{[\hat{\rho}_n, \hat{\Sigma}](\hat{g}^R \hat{\rho}_m - \hat{\rho}_m \hat{g}^A)\right\}. \quad (S17)$$

Since all our basis matrices $\hat{\rho}_n$ commute with both $\hat{\tau}_0$ and $\hat{\tau}_3$, we see that $V_{nm} = 0$ for a normal metal where $\hat{\Sigma} = \epsilon \hat{\tau}_3$. This implies that in the absence of other self-energy terms, all currents $I_n$ must be conserved in normal metals.

## C. Second-order self-energy terms

When describing phenomena such as spin-dependent scattering and orbital depairing [17], each self-energy contribution takes the form $\check{\Sigma} = \hat{\Sigma}\check{g}\hat{\Sigma}$. Substituting this into the commutator on the right-hand side of Eq. (S12), an explicit calculation yields

$$[\check{\Sigma}, \check{g}]^K = \hat{\Sigma}\hat{g}^R \hat{\Sigma}\hat{g}^K + \hat{\Sigma}\hat{g}^K \hat{\Sigma}\hat{g}^A - \hat{g}^R \hat{\Sigma}\hat{g}^K \hat{\Sigma} - \hat{g}^K \hat{\Sigma}\hat{g}^A \hat{\Sigma}. \quad (S18)$$

We recognize the right-hand side as a commutator with $\hat{\Sigma}$,

$$[\check{\Sigma}, \check{g}]^K = [\hat{\Sigma}, \hat{g}^R \hat{\Sigma}\hat{g}^K + \hat{g}^K \hat{\Sigma}\hat{g}^A]. \quad (S19)$$

We then multiply by $\hat{\rho}_n$ from the left and take the trace,

$$\mathrm{Tr}\left\{\hat{\rho}_n [\check{\Sigma}, \check{g}]^K\right\} = \mathrm{Tr}\left\{\hat{\rho}_n [\hat{\Sigma}, \hat{g}^R \hat{\Sigma}\hat{g}^K + \hat{g}^K \hat{\Sigma}\hat{g}^A]\right\}. \quad (S20)$$

Using the cyclic property of the trace, this can be rewritten as

$$\mathrm{Tr}\left\{\hat{\rho}_n [\check{\Sigma}, \check{g}]^K\right\} = \mathrm{Tr}\left\{[\hat{\rho}_n, \hat{\Sigma}](\hat{g}^R \hat{\Sigma}\hat{g}^K + \hat{g}^K \hat{\Sigma}\hat{g}^A)\right\}. \quad (S21)$$

Substituting in $\hat{g}^K = \hat{g}^R \hat{h} - \hat{h}\hat{g}^A$, the right-hand side becomes

$$\mathrm{Tr}\left\{[\hat{\rho}_n, \hat{\Sigma}](\hat{g}^R \hat{\Sigma}\hat{g}^R \hat{h} - \hat{h}\hat{g}^A \hat{\Sigma}\hat{g}^A + \hat{g}^R [\hat{h}, \hat{\Sigma}]\hat{g}^A)\right\}. \quad (S22)$$

Substituting the parametrization $\hat{h} = h_m \hat{\rho}_m$ into the above, and substituting the result back into Eq. (S12), we find that

$$\nabla \cdot I_n = -W_{nm} h_m, \quad (S23)$$

where we have defined the new quantity

$$W_{nm} \equiv \frac{i}{4} \mathrm{Tr}\left\{[\hat{\rho}_n, \hat{\Sigma}](\hat{g}^R \hat{\Sigma}\hat{g}^R \hat{\rho}_m - \hat{\rho}_m \hat{g}^A \hat{\Sigma}\hat{g}^A + \hat{g}^R [\hat{\rho}_m, \hat{\Sigma}]\hat{g}^A)\right\}. \quad (S24)$$

## D. Deriving the kinetic equation

In the previous subsections, we have shown that for a system described by a general second-order self-energy matrix $\check{\Sigma}$, which has contributions of the types $\check{\Sigma} = \hat{\Sigma}^{(1)}$ and $\check{\Sigma} = \hat{\Sigma}^{(2)}\check{g}\hat{\Sigma}^{(2)}$, the nonequilibrium distribution function satisfies the equations

$$\nabla \cdot \boldsymbol{I}_n = -(V_{nm} + W_{nm})h_m, \quad \text{(S25)}$$
$$\boldsymbol{I}_n = \boldsymbol{Q}_{nm}h_m + M_{nm}\nabla h_m. \quad \text{(S26)}$$

Combining these equations, we find a new differential equation for the distribution function components $h_m$:

$$M_{nm}\nabla^2 h_m = -(\nabla M_{nm} + \boldsymbol{Q}_{nm}) \cdot \nabla h_m \\ - (\nabla \cdot \boldsymbol{Q}_{nm} + V_{nm} + W_{nm})h_m. \quad \text{(S27)}$$

This is an *explicit linear differential equation* for the distribution function $h_m$. This can be made manifest by first multiplying by the $8 \times 8$ matrix $M^{-1}$ from the left, and then rewriting the equation in terms of a 16-element state vector $(h, \nabla h)$.

Note that all coefficient matrices depend only on the equilibrium solution, and can be precalculated *before* solving the kinetic equation. The coefficients do however depend on position, since the equilibrium propagators and self-energy terms may do so. In practice, one might therefore wish to precalculate the Jacobian of the differential equation at the discrete positions where the equilibrium problem was solved, and then interpolate between these when solving the kinetic equation. In our experience, linear interpolation may lead to convergence issues, while *e.g.* Catmull–Rom cubic splines work very well [19].

## III. BOUNDARY CONDITIONS

In order to solve Eq. (S27), we also need boundary conditions. In some cases, a satisfactory approximation can be obtained using transparent boundary conditions for the propagator $\check{g}$. The corresponding boundary conditions for the distribution are then trivial to obtain: $h_n$ is equal on both sides of the interface.

For more realistic interfaces, the boundary conditions are often written in terms of the matrix current $\check{I}$ that is flowing *outwards* from the interface. This directionality means that one typically has to flip the sign of the boundary condition at one end of a material, where the current is directed opposite from the coordinate axis. Furthermore, let us assume that this matrix current is a linear function of the distribution $\hat{h}$; we will later prove that this is the case for spin-active tunneling or reflecting interfaces. Denoting the distribution function on "this" side of the interface as $h_m$, and on the "other" side as $\underline{h}_m$, we get

$$I_n = \underline{C}_{nm}\underline{h}_m - C_{nm}h_m. \quad \text{(S28)}$$

Extracting the current component $I_n$ flowing out of the interface from Eq. (S8), and denoting the normal derivatives by $\nabla \to \partial$,

$$M_{nm}\partial h_m + (Q_{nm} + C_{nm})h_m = \underline{C}_{nm}\underline{h}_m. \quad \text{(S29)}$$

If one uses a numerical solver that minimizes interface residuals, this is a very suitable form of the equation; but if one requires an explicit form, the derivative $\partial h_m$ is also easy to isolate. Note that the coefficients should only depend on the equilibrium properties of the system, and can therefore be precalculated.

In the following derivations, we will use the notations

$$C_{nm} \equiv T_{nm} + R_{nm}, \quad \text{(S30)}$$
$$\underline{C}_{nm} \equiv \underline{T}_{nm}, \quad \text{(S31)}$$

where the symbols $T$ and $R$ refer to the boundary condition contributions from tunneling and reflection terms, respectively.

### A. Spin-dependent tunneling contributions

We will now derive kinetic boundary conditions for magnetic interfaces with spin-dependent tunneling. To leading order in the tunneling probability, and all orders in the polarization, the matrix current at such an interface can be written [20–23]:

$$2L\check{I} = Dt[F(\underline{\check{g}}), \check{g}]. \quad \text{(S32)}$$

Here, $\check{g}$ refers to the propagator on "this" side of the interface, $\underline{\check{g}}$ to the "other" side, and the matrix function $F$ is defined as

$$F(\underline{\check{g}}) = \underline{\check{g}} + \frac{P}{1+\sqrt{1-P^2}}\{\underline{\check{g}}, \hat{m}\} + \frac{1-\sqrt{1-P^2}}{1+\sqrt{1-P^2}}\hat{m}\underline{\check{g}}\hat{m}. \quad \text{(S33)}$$

The remaining parameters are the ratio $t \equiv G_T/G_0$ between tunneling conductance and bulk conductance, material length $L$, interface polarization $P$, and magnetization matrix $\hat{m} \equiv \boldsymbol{m} \cdot \hat{\boldsymbol{\sigma}}$, where $\boldsymbol{m}$ is a unit vector pointing along the average interface magnetization. Note that for unpolarized interfaces, we get $F(\underline{\check{g}}) = \underline{\check{g}}$, which simplifies the boundary condition above and the results below. For vacuum interfaces, we can also set $\underline{\check{g}} = 0$.

We start our derivation by noting that since $\hat{m}$ is diagonal in Keldysh space, the function $F$ has the following property:

$$F(\check{g})^{R,K,A} = F(\hat{g}^{R,K,A}). \quad \text{(S34)}$$

Applied to the commutator in Eq. (S32), we then get

$$\hat{I}^K = \frac{Dt}{2L}\left[F(\underline{\hat{g}}^R)\hat{g}^K + F(\underline{\hat{g}}^K)\hat{g}^A - \hat{g}^R F(\underline{\hat{g}}^K) - \hat{g}^K F(\underline{\hat{g}}^A)\right]. \quad \text{(S35)}$$

Substituting in $\hat{g}^K = \hat{g}^R \hat{h} - \hat{h}\hat{g}^A$, and grouping similar terms,

$$\hat{I}^K = \frac{Dt}{2L}\left[F(\underline{\hat{g}}^R)(\hat{g}^R \hat{h} - \hat{h}\hat{g}^A) - (\hat{g}^R \hat{h} - \hat{h}\hat{g}^A)F(\underline{\hat{g}}^A)\right] \\ + \frac{Dt}{2L}\left[F(\underline{\hat{g}}^R\underline{\hat{h}} - \underline{\hat{h}}\underline{\hat{g}}^A)\hat{g}^A - \hat{g}^R F(\underline{\hat{g}}^R\underline{\hat{h}} - \underline{\hat{h}}\underline{\hat{g}}^A)\right]. \quad \text{(S36)}$$

We then substitute in $\hat{h} = h_m\hat{\rho}_m$ and $\underline{\hat{h}} = \underline{h}_m\hat{\rho}_m$, multiply by $\hat{\rho}_n/4$ from the left, and take the trace. This results in a linear boundary condition $I_n = \underline{T}_{nm}\underline{h}_m - T_{nm}h_m$, where we identify

$$T_{nm} \equiv \frac{Dt}{8L}\text{Tr}\left\{\hat{\rho}_n\left[(\hat{g}^R\hat{\rho}_m - \hat{\rho}_m\hat{g}^A)F(\underline{\hat{g}}^A) \\ - F(\underline{\hat{g}}^R)(\hat{g}^R\hat{\rho}_m - \hat{\rho}_m\hat{g}^A)\right]\right\}, \quad \text{(S37)}$$

$$\underline{T}_{nm} \equiv \frac{Dt}{8L}\text{Tr}\left\{\hat{\rho}_n\left[F(\underline{\hat{g}}^R\hat{\rho}_m - \hat{\rho}_m\underline{\hat{g}}^A)\hat{g}^A \\ - \hat{g}^R F(\underline{\hat{g}}^R\hat{\rho}_m - \hat{\rho}_m\underline{\hat{g}}^A)\right]\right\}. \quad \text{(S38)}$$

Finally, using the cyclic trace rule, these results simplify to:

$$T_{nm} = \frac{Dt}{8L} \text{Tr}\left\{[F(\underline{\hat{g}}^A)\hat{\rho}_n - \hat{\rho}_n F(\underline{\hat{g}}^R)](\hat{g}^R\hat{\rho}_m - \hat{\rho}_m\hat{g}^A)\right\}, \quad \text{(S39)}$$

$$\underline{T}_{nm} = \frac{Dt}{8L} \text{Tr}\left\{(\hat{g}^A\hat{\rho}_n - \hat{\rho}_n\hat{g}^R)[F(\underline{\hat{g}}^R\hat{\rho}_m - \hat{\rho}_m\underline{\hat{g}}^A)]\right\}. \quad \text{(S40)}$$

### B. Spin-dependent reflection contributions

We will now derive the boundary coefficients for a spin-mixing interface. These boundary conditions can either be used alone, in the case of completely opaque interfaces to ferromagnetic insulators, or together with the tunneling boundary conditions from the previous subsection. The spin-mixing contribution to the matrix current is [21–25]:

$$2L\check{I} = -iDr[\hat{m}', \check{g}], \quad \text{(S41)}$$

where $r \equiv G_\varphi/G_0$ is the ratio between the spin-mixing and bulk conductances, and $\hat{m}' \equiv \bm{m}' \cdot \hat{\bm{\sigma}}$ is the interface magnetization matrix. In the case of inhomogeneous magnetic interfaces, $\bm{m}'$ may be different from $\bm{m}$, due to reflected and transmitted quasiparticles experiencing different average magnetizations.

Extracting the Keldysh component of the boundary condition, and substituting in $\hat{g}^K = \hat{g}^R\hat{h} - \hat{h}\hat{g}^A$ on the right-hand side,

$$\hat{I}^K = -\frac{iDr}{2L}[\hat{m}', \hat{g}^R\hat{h} - \hat{h}\hat{g}^A]. \quad \text{(S42)}$$

Substituting in $\hat{h} = h_m\hat{\rho}_m$, multiplying by $\hat{\rho}_n/4$ from the left, and taking the trace, we find the current components

$$I_n = -\frac{iDr}{8L} \text{Tr}\left\{\hat{\rho}_n[\hat{m}', \hat{g}^R\hat{\rho}_m - \hat{\rho}_m\hat{g}^A]\right\}h_m. \quad \text{(S43)}$$

Rewriting the commutator with the cyclic trace rule, and identifying the trace as a boundary coefficient, we conclude that the equation follows the pattern $I_n = -R_{nm}h_m$, where

$$R_{nm} \equiv -\frac{iDr}{8L} \text{Tr}\left\{[\hat{m}', \hat{\rho}_n](\hat{g}^R\hat{\rho}_m - \hat{\rho}_m\hat{g}^A)\right\}. \quad \text{(S44)}$$

### C. Nonequilibrium reservoirs

The boundary conditions above require knowledge of the distributions $\underline{h}$ in any reservoirs that couple to the system. By a *reservoir*, we mean a bulk material with a homogeneous quasiparticle distribution, which may be either in or out of equilibrium. In equilibrium, the electron density $n_e = \langle \Psi^\dagger\Psi \rangle$ should be described by Fermi–Dirac statistics $f(\epsilon) = 1/[1+\exp(\epsilon/T)]$, and the holes $n_h = \langle \Psi\Psi^\dagger \rangle$ by the remaining probability $1 - f(\epsilon)$, where the quasiparticle energy $\epsilon$ is measured relative to the Fermi level. This can be used to derive that the distribution is simply given by $\hat{h} = [1 - 2f(\epsilon)]\hat{\rho}_0$ in equilibrium, which reproduces the conventional expression $\hat{h} = \tanh(\epsilon/2T)\hat{\rho}_0$ [16, 26].

Upon applying a voltage $V$, the chemical potential of the reservoir is shifted by $eV$. This increases the electron density but decreases the hole density, and thus shifts the electron and hole blocks of the distribution above in opposite directions [16, 27]:

$$\hat{h} = \begin{pmatrix} \tanh[(\epsilon+eV)/2T]\sigma_0 & 0 \\ 0 & \tanh[(\epsilon-eV)/2T]\sigma_0 \end{pmatrix}. \quad \text{(S45)}$$

Taking the appropriate traces using Eq. (S5), one finds an energy mode $h_0$ and charge mode $h_4$, while the spin-energy modes $h_1, h_2, h_3$ and spin modes $h_5, h_6, h_7$ remain zero.

In a more general spin-dependent reservoir, the distribution matrix $\hat{h}$ should contain components proportional to $\sigma_1, \sigma_2, \sigma_3$ as well. One way to describe such a spin dependence, is that spin-up and spin-down particles experience different voltages $V_\uparrow$ and $V_\downarrow$, and possibly different temperatures $T_\uparrow$ and $T_\downarrow$ [18, 28]. Physically, the most extreme realization of this situation is given by half-metallic reservoirs, which only have one metallic spin-band that can couple to regular conductors [21, 29]. If we for simplicity take the spin-quantization axis to be the $z$-axis, introducing spin-dependent voltages and temperatures yields

$$\hat{h} = \begin{pmatrix} \tanh[(\epsilon+eV_\uparrow)/2T_\uparrow] & 0 & 0 & 0 \\ 0 & \tanh[(\epsilon+eV_\downarrow)/2T_\downarrow] & 0 & 0 \\ 0 & 0 & \tanh[(\epsilon-eV_\uparrow)/2T_\uparrow] & 0 \\ 0 & 0 & 0 & \tanh[(\epsilon-eV_\downarrow)/2T_\downarrow] \end{pmatrix}.$$

This can also be parametrized in terms of an average voltage $V \equiv (V_\uparrow + V_\downarrow)/2$ and spin-voltage $V_s \equiv (V_\uparrow - V_\downarrow)/2$; in non-magnetic materials, a gradient in the former gives rise to a pure electric current, and the latter a pure spin current. Similarly, one can define an average temperature $T \equiv (T_\uparrow + T_\downarrow)/2$ and spin-temperature $T_s \equiv (T_\uparrow - T_\downarrow)/2$, whose gradients cause energy and spin-energy currents. Finally, the physics of the system should not depend on our arbitrary choice of coordinate axes, so a corresponding expression for a general spin quantization axis $\bm{u} = (u_1, u_2, u_3)$ can be obtained using spin rotation matrices.

Introducing spin-voltages and spin-temperatures as discussed above, performing a spin rotation to an arbitrary spin axis $\bm{u}$, and again applying Eq. (S5), we find the general result:

$$\begin{aligned}
h_0 &= [h_{++} + h_{+-} + h_{-+} + h_{--}], \\
h_1 &= [h_{++} - h_{+-} + h_{-+} - h_{--}]u_1, \\
h_2 &= [h_{++} - h_{+-} + h_{-+} - h_{--}]u_2, \\
h_3 &= [h_{++} - h_{+-} + h_{-+} - h_{--}]u_3, \\
h_4 &= [h_{++} + h_{+-} - h_{-+} - h_{--}], \\
h_5 &= [h_{++} - h_{+-} - h_{-+} + h_{--}]u_1, \\
h_6 &= [h_{++} - h_{+-} - h_{-+} + h_{--}]u_2, \\
h_7 &= [h_{++} - h_{+-} - h_{-+} + h_{--}]u_3.
\end{aligned} \quad \text{(S46)}$$

For brevity, the above is written in terms of the distributions

$$h_{cs} = \tanh[(\epsilon + cV + csV_s)/(T + sT_s)]/4, \quad \text{(S47)}$$

which describe quasiparticles with a charge index $c$ and spin index $s$. For instance, $h_{+-}$ corresponds to $c = +1$ (electrons) and $s = -1$ (spin-down), and so on. This describes a quite general reservoir that can have a voltage, spin-voltage, temperature, and spin-temperature, with an arbitrary spin-quantization axis. The main possibility not accounted for, is that the spin-splitting of the voltage and temperature could in principle be in different directions. Also, it might be possible to excite some even more exotic distributions via *e.g.* optical methods, which might be unnatural to describe in terms of voltages and temperatures.

## IV. NUMERICAL RESULTS FOR THE DISTRIBUTION

In this section, we focus on how the numerically calculated distribution function $\hat{h}$ varies as a function of position $x$ and energy $\epsilon$. To briefly reiterate, the system we consider consists of a conventional superconductor of length $L = 8\xi$, connected to normal contacts at voltages $\pm V/2$ via tunneling interfaces. Since a full decomposition and exposition of the distribution function takes a lot of space to visualize, we focus on the parameters $m = eV/2 = \Delta_0$ at a low temperature $T = 0.01 T_c$. We note that the results for other magnetic fields and voltages are qualitatively similar to the ones shown here, and the results at higher temperatures are basically just thermally smeared out.

Following the notation of previous sections, we parametrize the distribution function in terms of the modes $h_n = \text{Tr}[\hat{\rho}_n \hat{h}]/4$. For a system with a homogeneous magnetic field along the $z$-axis, only four components may be nonzero: the energy mode $h_0$, spin-energy mode $h_3$, charge mode $h_4$, and pure spin mode $h_7$. Numerically, these have only been calculated for positive energies $\epsilon > 0$. However, the energy and spin-energy modes are by definition odd functions of $\epsilon$, while the charge and spin modes are even functions of $\epsilon$, and these energy symmetries can be used to find obtain the distribution at $\epsilon < 0$.

The results are shown in Fig. S2. We see that the energy mode $h_0$ agrees perfectly with the analytically expected result

$$h_0 = \frac{1}{2}\{\tanh[(\epsilon + eV/2)/2T] + \tanh[(\epsilon - eV/2)/2T]\}, \quad \text{(S48)}$$

which at low temperatures gives $h_0 = 0$ for $0 < \epsilon < eV/2$ and $h_0 = 1$ for $\epsilon > eV/2$, where $eV/2 = \Delta_0$ here. In general, an energy-mode excitation relaxes over the inelastic scattering length, which we take to be long compared to the system size. We note that in the most relevant temperature range for experiments (below ∼ 1 K), electron–electron interactions are the dominant contribution to the inelastic scattering length [30], which appears to diverge at lower temperatures [31].

The charge mode is an order of magnitude smaller than the energy mode at the interfaces. Since the voltages $\pm V/2$ at the interfaces are opposite, the charge mode is also forced to be an antisymmetric function of position. In total, the charge mode is therefore much smaller than the energy mode even at the interfaces, and vanishes completely deeper inside the superconductor. This helps to explain the remarkably accurate agreement between the analytical and numerical calculations presented in the main manuscript [e.g. in Fig. 2(a) and 3(a)], and legitimizes the approximation $\hat{h} \approx h_0 \rho_0$ in the superconductor.

In addition to the energy and charge modes, which were explained in the main manuscript, we see that there is also a small spin-energy and pure spin mode in the system. However, these are roughly two orders of magnitude smaller than the energy mode, which explains why these are not essential for the analytical understanding presented in the main manuscript.

The origin of the spin-energy mode is actually straightforward to understand. As can be read out from Fig. 3(a) of the main manuscript, the self-consistent order parameter $\Delta \approx \Delta_0/4$ for $m = eV/2 = \Delta_0$. Fig. 3(b) and related discussion shows that this causes two gaps in the DOS: a hard gap in the spin-down DOS centered at $\epsilon = +m = +\Delta_0$, and a hard gap in the spin-up DOS centered at $\epsilon = -m = -\Delta_0$. This spin-split superconducting state is then coupled to a reservoir at a voltage $eV/2 = +\Delta_0$ at the left end $x = 0$. There is therefore a region $(3/4)\Delta_0 < \epsilon < \Delta_0$ where spin-up quasiparticles are injected while the spin-down band has a hard gap. At the right end $x = 8\xi$, the material similarly couples to a reservoir at a voltage $-eV/2 = -\Delta_0$. This causes spin-down quasiparticles to be drained from the region $-\Delta_0 < \epsilon < -(3/4)\Delta_0$, while the spin-up band has a hard gap. Thus, one has effectively injected spin-up electrons for $(3/4)\Delta_0 < \epsilon < \Delta_0$ and injected spin-down holes for $-\Delta_0 < \epsilon < -(3/4)\Delta_0$, which is an excitation of the spin-energy mode. Note that this energy region where the charge mode of the voltage-biased reservoirs couple to a spin-energy mode in the superconductor, is the same region where the charge mode in the superconductor is slightly weakened.

Finally, in Fig. S3, we show how the spin-independent occupation number $n(\epsilon) = [1 - h_0(\epsilon) - h_4(\epsilon)]/2$ depends on position. This has a two-step Fermi-Dirac-like shape throughout the superconductor, but has some variation through the superconductor due to the varying charge mode.

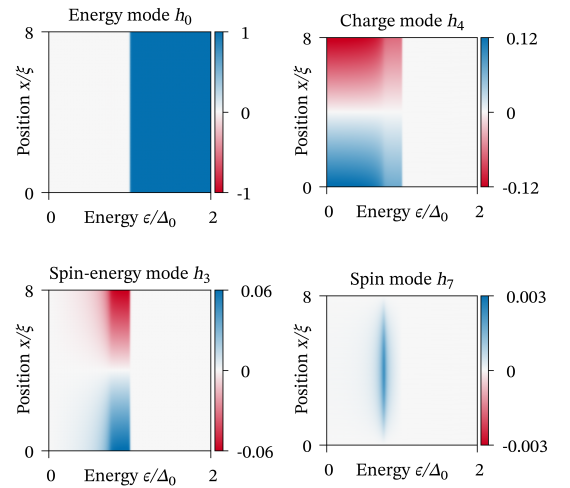

FIG. S2: Nonequilibrium modes $h_n$ of the distribution function $\hat{h}$ as function of position and energy. Note that the colorbars used for the different panels differ by several orders of magnitude.

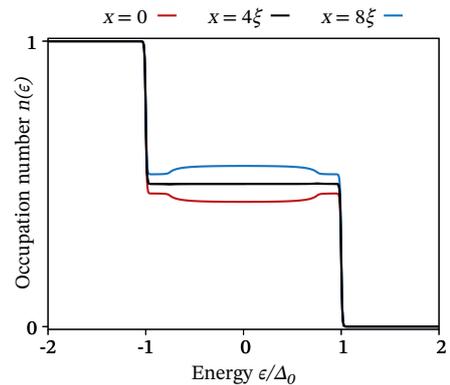

FIG. S3: Spin-independent occupation number $n(\epsilon)$ at different positions in the superconductor. The distribution has a two-step shape throughout the material, but a local charge accumulation distorts it slightly near the interfaces to the reservoirs.

## V. CONVERSION TO SUPERCURRENTS

For the numerical calculations presented in the main manuscript, the interfaces between the superconductor and voltage-biased reservoirs were assumed to be relatively opaque. However, we also performed numerical tests for other parameter sets, including voltage-biased N/S/N junctions with completely transparent interfaces. These showed that superconductivity remained stable in high magnetic fields $m$ for a voltage bias $eV/2 = m$, provided that the superconductor is sufficiently long compared with the coherence length $\xi$. The fact that a voltage drop can exist across a "superconductor" even in the absence of interface resistance may seem a bit surprising. The answer has previously been derived in e.g. Refs. 15 and 32. When a resistive current is injected into a superconductor via ideal interfaces, there is actually a layer of thickness $\sim \xi$ where the resistive current is converted into a supercurrent. In other words, there is still a "superconducting contact resistance" in the transparent limit, and the associated voltage drop occurs near the interfaces. For more details, see e.g. the numerical results in Ref. 32.

In a superconductor with a spin-splitting that exceeds the order parameter, an even more peculiar situation arises. Since the DOS is no longer gapped at the Fermi level, and resistive currents are only converted into supercurrents in the gapped regions of the energy spectrum, a long-ranged resistive current can exist in the superconductor in this limit. On the following pages, we derive an approximate analytical result for the length scale over which resistive currents decay inside strongly spin-split superconductors to explain this observation. The derivation itself makes a number of approximations, some more reasonable than others, but the final analytical equation is simple and agrees quite well with our numerical observations.

We should stress that even though the superconductor can harbour a long-ranged resistive current in this exotic state, the name "superconductor" is still fitting. The most fundamental way to justify it is that the material still exhibits a singlet order parameter $\Delta$ and a spontaneously broken $U(1)$ symmetry, which are the hallmarks of a superconducting state. Another perspective is that the material also supports dissipationless currents when the order parameter has a phase-winding. This would perhaps be even clearer if we used the experimental setup sketched in Fig. 1(b) of the main manuscript. There, no resistive current is injected into the superconductor, and yet the same stabilization effect at $m = eV/2$ occurs. If a supercurrent is then generated using e.g. a weak out-of-plane magnetic field, where the flux couples directly to the phase, we would generate a pure supercurrent in the system. The conclusion is that the ability of a material to host resistive and dissipationless currents over long distances are not always mutually exclusive.

The starting point for our analytical derivation is Eq. (S27). Let us assume that the system under consideration is roughly homogeneous, so that $\nabla g^R \approx \nabla g^A \approx 0$, in which case $\boldsymbol{Q} \approx 0$ and $\nabla M \approx 0$. Furthermore, let us assume that no spin-flip scattering, spin-orbit scattering, or orbital depairing effects are important in the system, so that the term $W = 0$ as well. The kinetic equation then reduces to the much simpler form

$$M_{nm}\nabla^2 h_m = -V_{nm}h_m. \tag{S49}$$

In the most general case, $M$ and $V$ are $8 \times 8$ matrices while $h$ is an 8-vector. If one considers a system where only one spin axis is relevant, such as a bulk superconductor with a spin-splitting along the $z$-axis, this can be reduced to a system of $4 \times 4$ matrices in the equation for a 4-vector $h$. This gives us a system of coupled equations for the charge, energy, spin, and spin-energy modes of the nonequilibrium distribution function, which can in principle be solved explicitly.

We will now assume that the dominant relaxation process of the charge mode inside a superconductor occurs via the diagonal terms. The charge mode is given by $h_4$ in our notation, so neglecting the coupling to other modes, we then get

$$\nabla^2 h_4 \approx -(V_{44}/M_{44})h_4. \tag{S50}$$

Comparing this to the equation $\nabla^2 h_4 = h_4/\lambda^2$ that would define an exponential decay, we can describe such a decay via an energy-dependent *charge relaxation length* $\lambda \equiv \sqrt{-M_{44}/V_{44}}$.

The source term $V_{44}$ above can be calculated using Eq. (S17),

$$V_{nm} = \frac{i}{4}\text{Tr}\left\{[\hat{\rho}_n, \hat{\Sigma}](\hat{g}^R\hat{\rho}_m - \hat{\rho}_m\hat{g}^A)\right\}. \tag{S51}$$

We are interested in the case $n = m = 4$, and since the basis matrix $\hat{\rho}_4 = \hat{\tau}_3\sigma_0$, the commutator $[\hat{\rho}_4, \hat{\Sigma}]$ vanishes for the self-energy terms corresponding to a regular magnet $\hat{\Sigma} = \epsilon\hat{\tau}_3 + m\hat{\sigma}_3$. The charge mode is therefore controlled by the superconducting contributions $\hat{\Sigma} = \hat{\Delta} = \text{antidiag}(+\Delta, -\Delta, +\Delta^*, -\Delta^*)$. If we choose a real gauge, which is possible since we already assumed that any supercurrents are negligible $\boldsymbol{Q} \approx 0$, this reduces to $\hat{\Sigma} = \Delta i\hat{\tau}_1\sigma_2$. Substituted into the equation above, we find that

$$V_{44} = -\frac{\Delta}{4}\text{Tr}\left\{[\hat{\tau}_3, \hat{\tau}_1\sigma_2](\hat{g}^R\hat{\tau}_3 - \hat{\tau}_3\hat{g}^A)\right\}. \tag{S52}$$

Using $[\hat{\tau}_3, \hat{\tau}_1] = 2i\hat{\tau}_2$ and $\hat{g}^A = -\hat{\tau}_3\hat{g}^{R\dagger}\hat{\tau}_3$, this becomes

$$V_{44} = -\frac{i\Delta}{2}\text{Tr}\left\{\hat{\tau}_2\sigma_2(\hat{g}^R\hat{\tau}_3 + \hat{g}^{R\dagger}\hat{\tau}_3)\right\}. \tag{S53}$$

We then use the cyclic rule to move the $\hat{\tau}_3$ matrices to the other end of the trace, and use the Pauli identity $\hat{\tau}_3\hat{\tau}_2 = -i\hat{\tau}_1$, and finally apply $\text{Tr}[A + A^\dagger] = 2\,\text{Re}\,\text{Tr}[A]$ to simplify the result:

$$V_{44} = -\Delta\,\text{Re}\,\text{Tr}\left\{\hat{\tau}_1\sigma_2\hat{g}^R\right\}. \tag{S54}$$

This is essentially the same result as was used in Ref. 15 to show that resistive currents decay over a length $\sim \xi$ inside a superconductor, except that we will attempt to use it for the more general case of a spin-split superconductor.

In general, we can write the retarded propagator as

$$\hat{g}^R = \begin{pmatrix} +g & +f \\ -\tilde{f} & -\tilde{g} \end{pmatrix}. \tag{S55}$$

Multiplying this by $\hat{\tau}_1\sigma_2$ and taking the trace, we find that

$$\text{Tr}\left\{\hat{\tau}_1\sigma_2\hat{g}^R\right\} = \text{Tr}\left\{\sigma_2(f - \tilde{f})\right\}. \tag{S56}$$

Firstly, we can split the anomalous propagators into singlets and triplets using the decomposition $f = (f_s + \boldsymbol{f}_t \cdot \boldsymbol{\sigma})i\sigma_2$. Secondly, when the superconducting gap $\Delta$ is purely real, the singlet component satisfies $\tilde{f}_s = -f_s$. This leads us to the conclusion that the only contribution to the trace of $\sigma_2(f - \tilde{f})$ comes from the singlet part $\sigma_2(f_s + f_s)i\sigma_2 = 2if_s\sigma_0$:

$$\text{Tr}\{\sigma_2(f - \tilde{f})\} = 4if_s. \tag{S57}$$

Thus, the final form of the source term $V_{44}$ derived above is

$$V_{44} = 4\Delta \, \text{Im}(f_s), \tag{S58}$$

where as usual the singlet component $f_s \equiv (f_{\uparrow\downarrow} - f_{\downarrow\uparrow})/2$.

We now turn to the matrix $M$, which can be interpreted as an energy-dependent renormalized diffusion coefficient [16]. In previous sections, this quantity was defined as

$$M_{nm} = \frac{D}{4}\text{Tr}\{\hat{\rho}_n\hat{\rho}_m - \hat{\rho}_n\hat{g}^R\hat{\rho}_m\hat{g}^A\}. \tag{S59}$$

We again set $n = m = 4$, and use $\hat{\rho}_4 = \hat{\tau}_3\sigma_0$, $\hat{g}^A = -\hat{\tau}_3\hat{g}^{R\dagger}\hat{\tau}_3$,

$$M_{44} = \frac{D}{4}\text{Tr}\{\hat{\tau}_3\hat{\tau}_3 + \hat{\tau}_3\hat{g}^R\hat{g}^{R\dagger}\hat{\tau}_3\}. \tag{S60}$$

Applying the cyclic trace rule, and the identity $\hat{\tau}_3^2 = 1$, we get

$$M_{44} = \frac{D}{4}\text{Tr}\{1 + \hat{g}^R\hat{g}^{R\dagger}\}. \tag{S61}$$

One way to parametrize the retarded propagator $\hat{g}^R$ is [33]

$$\hat{g}^R = \begin{pmatrix} (g_s + \boldsymbol{g}_t \cdot \boldsymbol{\sigma}) & (f_s + \boldsymbol{f}_t \cdot \boldsymbol{\sigma})i\sigma_2 \\ -i\sigma_2(\tilde{f}_s - \tilde{\boldsymbol{f}}_t \cdot \boldsymbol{\sigma}) & -\sigma_2(\tilde{g}_s - \tilde{\boldsymbol{g}}_t \cdot \boldsymbol{\sigma})\sigma_2 \end{pmatrix}. \tag{S62}$$

Explicitly taking the complex-transpose of this matrix, and using that $\sigma_n$ are Hermitian while $i$ is anti-Hermitian, we find

$$\hat{g}^{R\dagger} = \begin{pmatrix} (g_s^* + \boldsymbol{g}_t^* \cdot \boldsymbol{\sigma}) & (\tilde{f}_s^* - \tilde{\boldsymbol{f}}_t^* \cdot \boldsymbol{\sigma})i\sigma_2 \\ -i\sigma_2(f_s^* + \boldsymbol{f}_t^* \cdot \boldsymbol{\sigma}) & -\sigma_2(\tilde{g}_s^* - \tilde{\boldsymbol{g}}_t^* \cdot \boldsymbol{\sigma})\sigma_2 \end{pmatrix}. \tag{S63}$$

We now calculate the product $\hat{g}^R\hat{g}^{R\dagger}$, keeping only diagonal terms proportional to an even power of Pauli matrices, since terms proportional to $\boldsymbol{\sigma}$ disappear when we later take the trace:

$$\hat{g}^R\hat{g}^{R\dagger} = \begin{pmatrix} |g_s|^2 + |\boldsymbol{g}_t|^2 + |f_s|^2 + |\boldsymbol{f}_t|^2 & \cdots \\ \cdots & |\tilde{f}_s|^2 + |\tilde{\boldsymbol{f}}_t|^2 + |\tilde{g}_s|^2 + |\tilde{\boldsymbol{g}}_t|^2 \end{pmatrix}. \tag{S64}$$

Due to the electron-hole symmetry of quasiclassical theory, it is reasonable to expect all *magnitudes* $|g_s|^2$, $|\boldsymbol{g}_t|^2$, $|f_s|^2$, $|\boldsymbol{f}_t|^2$ to be invariant under tilde-conjugation, even though the signs of the quantities themselves might change. Using this, we find

$$\text{Tr}\{1 + \hat{g}^R\hat{g}^{R\dagger}\} = 4 + 4|g_s|^2 + 4|\boldsymbol{g}_t|^2 + 4|f_s|^2 + 4|\boldsymbol{f}_t|^2. \tag{S65}$$

Going back to our result for $M_{44}$, we find the final result:

$$M_{44} = D(1 + |g_s|^2 + |\boldsymbol{g}_t|^2 + |f_s|^2 + |\boldsymbol{f}_t|^2). \tag{S66}$$

Putting together the pieces we have calculated so far, we find that the charge relaxation length $\lambda = \sqrt{-M_{44}/V_{44}}$ is:

$$\lambda = \sqrt{\frac{D(1 + |g_s|^2 + |\boldsymbol{g}_t|^2 + |f_s|^2 + |\boldsymbol{f}_t|^2)}{-4\Delta \, \text{Im} \, f_s}}. \tag{S67}$$

Defining the coherence length $\xi' \equiv \sqrt{D/\Delta}$, which depends on temperature via the self-consistent gap $\Delta$, this becomes

$$\lambda = \frac{\xi'}{2}\sqrt{\frac{1 + |g_s|^2 + |\boldsymbol{g}_t|^2 + |f_s|^2 + |\boldsymbol{f}_t|^2}{-\text{Im} \, f_s}}. \tag{S68}$$

This is a somewhat general result, as it is valid regardless what mixture of singlets and triplets is present in the system, and we have not made any assumptions of weak proximity or weak inverse proximity effect. The result does, however, rest on two crucial assumptions. The first is that the charge mode couples only weakly to the other nonequilibrium modes of the distribution function, so that it was sufficient to consider the diagonal parts of Eq. (S49). This should be a reasonable approximation as long as either (i) the charge mode relaxes over a shorter length scale than the other modes, or (ii) the coupling to the other modes is weak. The second assumption is that the system is roughly homogeneous, so that we can neglect variations in the propagator and the presence of any supercurrents. In practice, this should be a fair approximation if we consider a large superconductor with tunneling contacts.

Let us first consider the numerator of Eq. (S68). We see that the numerator is always larger than 1. Furthermore, for a normal metal $|g_s|^2 = 1$ while the other quantities are zero, making the numerator simply equal to 2. On the other hand, for a BCS superconductor, we get $|g_s|^2 = |\epsilon^2/(\epsilon^2 - \Delta^2)|$ while $|f_s|^2 = |\Delta^2/(\epsilon^2 - \Delta^2)|$. From this, we find that $|g_s|^2 + |f_s|^2 \approx 1$ in the limits $|\epsilon| \ll \Delta$ and $|\epsilon| \gg \Delta$, but diverges as $\epsilon \to \pm\Delta$. Thus, except near the coherence peaks of a superconductor, the charge relaxation length should mainly be controlled by the denominator in Eq. (S68), yielding the approximation

$$\lambda \approx \frac{\xi}{\sqrt{-2\,\text{Im}\,f_s}}. \tag{S69}$$

If we again focus on a BCS superconductor, $f_s = \Delta/\sqrt{\epsilon^2 - \Delta^2}$ for $\epsilon > 0$. For energies outside the gap $\epsilon > \Delta$, we see that $\text{Im}\,f_s = 0$, yielding a charge relaxation length $\lambda \to \infty$. On the other hand, for energies inside the gap $\epsilon \ll \Delta$, we find that $\text{Im}\,f_s = -1$, yielding $\lambda \approx \xi/\sqrt{2}$. Thus, we found exactly the kind of behaviour we were expecting: the charge mode is screened over a characteristic length $\sim \xi$ inside the gap, but is not screened for energies that reside outside the gap.

In Figs. S4 and S5, we show how the DOS $N(\epsilon)$ and approximate charge relaxation length $\lambda(\epsilon)$ vary with the spin-splitting $m$ in a superconductor. The results confirm that resistive currents decay over a length $\sim \xi$ in gapped parts of the spectrum, while a long-ranged resistive current can exist in ungapped parts of the spectrum. This result agrees quite well with our numerical results, where we observe that for $m = eV/2 > \Delta$, a resistive current contribution can persist throughout the superconductor for the energy range between the two spectral gaps.

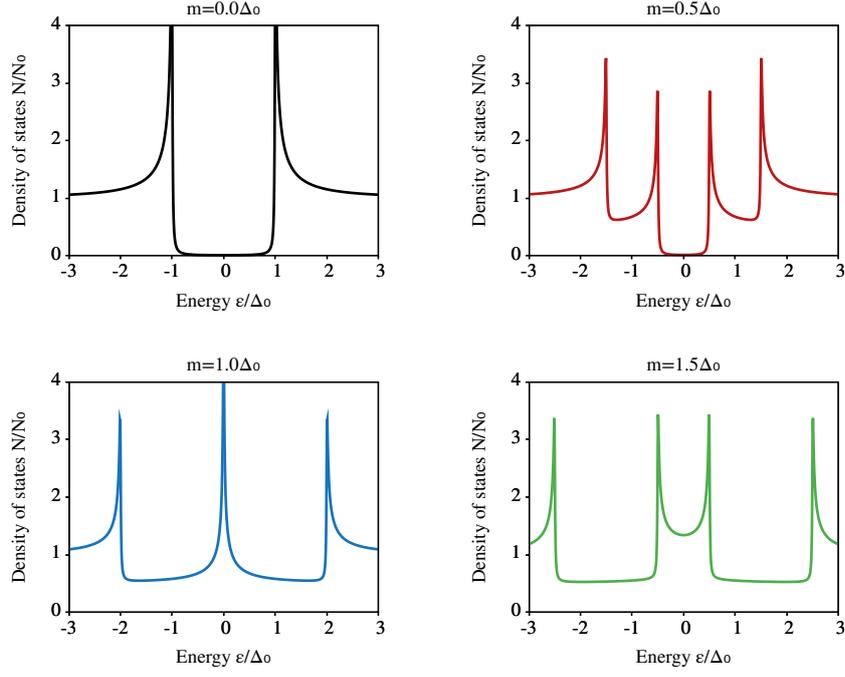

FIG. S4: DOS for a Zeeman-split superconductor with a magnetic field $m$. For simplicity, this was calculated using a non-selfconsistent analytical solution for a bulk superconductor with an exchange field, using an inelastic scattering parameter $\epsilon \to \epsilon + 0.01i\Delta_0$.

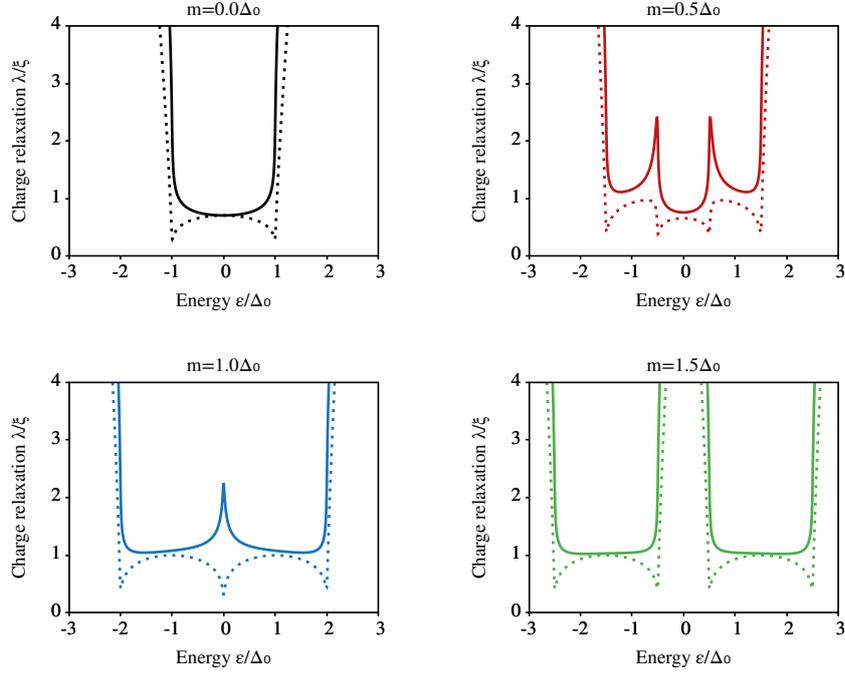

FIG. S5: Charge relaxation length for a Zeeman-split superconductor with a magnetic field $m$. The solid lines show the exact results from Eq. (S68), which we see align very well with how the DOS looks in Fig. S4. The dotted lines show the approximation in Eq. (S69), which manages to predict where the charge relaxation length is finite and infinite, but does not replicate its precise shape.